\documentclass[aps,prd,10pt,twocolumn,preprintnumbers,superscriptaddress,bibnotes,nofootinbib,letterpaper,longbibliography]{revtex4-2}

\usepackage{xcolor}
\usepackage{gensymb}
\definecolor{mygreen}{RGB}{0,108,101}
\definecolor{myorange}{RGB}{255, 128, 1}
\definecolor{myorange}{HTML}{FF7518}
\definecolor{osu}{HTML}{ba0c2f}
\definecolor{ccapp}{HTML}{8a8e3a}
\definecolor{buckeyeblue}{HTML}{001C5A}

\usepackage{graphicx}
\usepackage{amsmath, amssymb}
\usepackage{hyperref}
\usepackage{booktabs}
\hypersetup{
  colorlinks  = true,
  urlcolor    = osu,
  linkcolor   = blue,
  citecolor   = red,
}
\usepackage[capitalize]{cleveref}

\newcommand{\orcid}[1]{\begingroup
  \hypersetup{hidelinks}\href{https://orcid.org/#1}{\includegraphics[width=10pt]{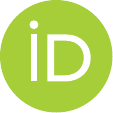}\,} \endgroup}

\begin{document}

\title{Self-Calibration of the Neutrino-Argon Cross Section with Solar Neutrinos}

\author{Rasmi Hajjar\,\orcid{0000-0002-9227-5364}}
\email{hajjar.44@osu.edu}
\affiliation{Center for Cosmology and AstroParticle Physics (CCAPP), \href{https://ror.org/00rs6vg23}{\color{osu}Ohio State University}, Columbus, OH 43210}
\affiliation{Department of Physics, \href{https://ror.org/00rs6vg23}{\color{osu}Ohio State University}, Columbus, OH 43210}
\affiliation{Department of Astronomy, \href{https://ror.org/00rs6vg23}{\color{osu}Ohio State University}, Columbus, OH 43210}

\author{Obada Nairat\,\orcid{0000-0003-2019-9021}}
\email{nairat.2@osu.edu}
\affiliation{Center for Cosmology and AstroParticle Physics (CCAPP), \href{https://ror.org/00rs6vg23}{\color{osu}Ohio State University}, Columbus, OH 43210}
\affiliation{Department of Physics, \href{https://ror.org/00rs6vg23}{\color{osu}Ohio State University}, Columbus, OH 43210}

\author{John F. Beacom\,\orcid{0000-0002-0005-2631}}
\email{beacom.7@osu.edu}
\affiliation{Center for Cosmology and AstroParticle Physics (CCAPP), \href{https://ror.org/00rs6vg23}{\color{osu}Ohio State University}, Columbus, OH 43210}
\affiliation{Department of Physics, \href{https://ror.org/00rs6vg23}{\color{osu}Ohio State University}, Columbus, OH 43210}
\affiliation{Department of Astronomy, \href{https://ror.org/00rs6vg23}{\color{osu}Ohio State University}, Columbus, OH 43210}

\date{\today}


\begin{abstract}
The success of DUNE's MeV physics program depends upon high-precision knowledge of the charged-current (CC) $\nu_e+\mathrm{^{40}Ar}$ cross section.  While there are indirect constraints at the 10\% level for the nuclear transitions that constitute this cross section, the only direct measurement in the MeV range has an uncertainty of $\sim$50\%. We show, surprisingly, that the cross section can be precisely measured using the solar-neutrino data themselves. This is possible because of independent knowledge of the $^8$B flux and survival probability, plus the distinctive angular distributions of the Fermi and Gamow-Teller transitions that comprise the cross section.  We propose new methods to extract the transition strengths, considering both intuitive groupings and a Principal Component Analysis.  Under pessimistic assumptions about detection, but taking detector uncertainties to be controlled, \emph{we demonstrate that a precision of $\lesssim$2\% on the cross section can be achieved in the 9--15~MeV energy range.}  These results will be an important foundation for studying the cross section up to several tens of MeV, where the complexity increases significantly due to nuclear breakup channels but where reducing uncertainties is critical for supernova and atmospheric neutrino studies.
\end{abstract}

\maketitle

\section{Introduction}\label{sec:intro}

The Deep Underground Neutrino Experiment (DUNE) will play a central role in shaping the future of neutrino physics in the coming decades~\cite{DUNE:2018tke, DUNE:2018hrq, DUNE:2018mlo, DUNE:2016rla}. Liquid argon detectors such as DUNE constitute a technological and scientific breakthrough, introducing three-dimensional imaging. DUNE’s primary goal is precision studies of GeV neutrinos from a long-baseline accelerator~\cite{DUNE:2020jqi, DUNE:2020lwj, DUNE:2020fgq, DUNE:2024wvj, DUNE:2025sjq}. It will make extremely precise measurements of neutrino mixing, including neutrino unitarity tests~\cite{Fukugita:1986hr, Nunokawa:2007qh, Davidson:2008bu, Parke:2015goa, Qian:2015waa}.

DUNE's unique capabilities also position it to become a frontrunner in MeV neutrino astrophysics, especially because of its special sensitivity to electron neutrinos. One of DUNE's official goals is to detect a Milky Way supernova neutrino burst~\cite{DUNE:2020zfm}.  Following the comprehensive proposals of Refs.~\cite{Capozzi:2018dat, Zhu:2018rwc, Meighen-Berger:2024xbx} that DUNE could also make world-leading measurements of solar neutrinos, the prospects look challenging but encouraging, perhaps requiring additional modules~\cite{Bahcall:1986ry, Parsa:2022mnj, MicroBooNE:2023sxs, Garcia-Peris:2025exn, RuizFerreira:2025xzg}.

\begin{figure}[t]
    \centering \includegraphics[width=\columnwidth]{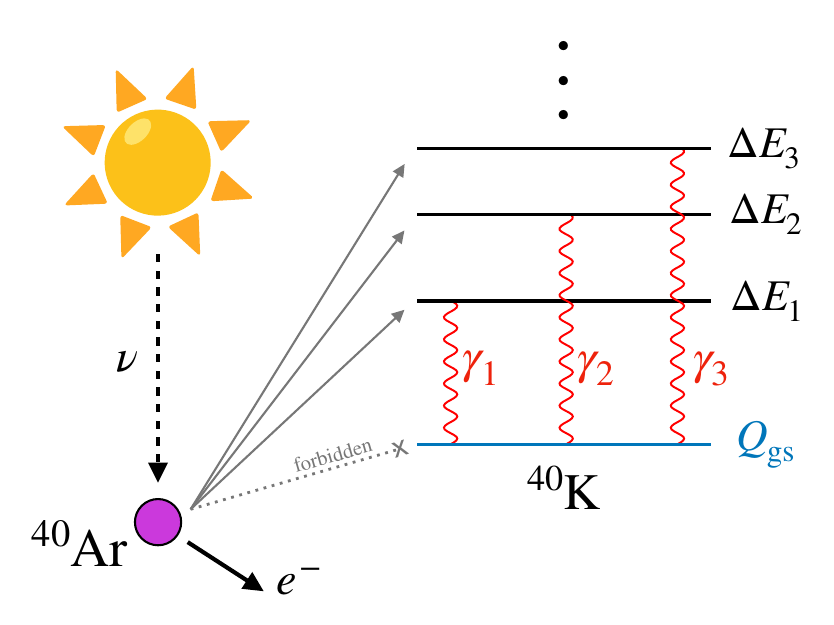}
    \caption{Schematic representation of a solar neutrino interacting with an argon nucleus, as in \cref{eq:nuAr_interaction,eq:K_deexcitation}. The final products of an interaction consist of an electron, a $\mathrm{^{40}K}$ nucleus, and de-excitation gamma rays  accounting for the transition energy, $\Delta E_i$.}
    \label{fig:nuAr_xsec}
\end{figure}

Figure~\ref{fig:nuAr_xsec} illustrates the steps in the dominant interaction, charged-current (CC) scattering~\cite{Raghavan:1986hv, Ormand:1994js, Trinder:1997xr, Bhattacharya:1998hc, ICARUS:1998nzl, Bhattacharya:2009zz, Capozzi:2018dat},
\begin{equation}\label{eq:nuAr_interaction}
\nu_e + \mathrm{^{40}Ar} \longrightarrow e^- + \mathrm{^{40}K}^*\,,
\end{equation}
where the * indicates nuclear excited states, as transitions to the ground state are forbidden. The final excited states promptly emit de-excitation gamma rays,
\begin{equation}\label{eq:K_deexcitation}
\mathrm{^{40}K}^* \longrightarrow \mathrm{^{40}K} + \gamma\,.
\end{equation}
A significant challenge to developing DUNE's low-energy astrophysical program is that the cross section is not adequately known.  The best measurements, which are indirect, are uncertain at the $\sim$10\% level~\cite{Capozzi:2018dat}, which is inadequate for discovery-level sensitivity in the MeV regime.

Very recently, a first measurement of the CC $\nu_e + \mathrm{^{40}Ar}$ cross section with solar neutrinos was made by the Dark Matter Experiment using Argon Pulseshape discrimination (DEAP) experiment~\cite{DEAP-3600:2017ker, DEAP:2026img}, which is a $\sim$3-ton liquid-argon detector primarily intended to probe dark matter.  By assuming standard solar fluxes and oscillation probabilities, they obtained an overall normalization for the cross section in the energy range 12.0–14.5~MeV.  They observed 6 counts, where the prediction is $2.3 \pm 0.4$ counts from solar neutrinos and $0.48^{+0.16}_{-0.15}$ counts from detector backgrounds.  From this, they derived that the cross section is $2.4^{+1.3}_{-1.0}$ times larger than expected from Ref.~\cite{Bhattacharya:2009zz}.  This only heightens the need for better measurements and our proposed approach.

In this paper, we demonstrate the capability of DUNE to extract the CC $\nu_e+\mathrm{^{40}Ar}$ cross section through the observation of $^8\mathrm{B}$ solar neutrinos. As we describe in main text, this is possible because of two key factors, hinted at in Ref.~\cite{Capozzi:2018dat}. First, the solar neutrino flux is exceptionally well characterized~\cite{Winter:2004kf, SNO:2011hxd} and the electron survival probability is well known (and nearly constant)~\cite{SNO:2011hxd, BOREXINO:2018ohr, Super-Kamiokande:2023jbt} in the energy range of interest. Second, the angular distributions of the outgoing electrons from Fermi and Gamow-Teller (GT) transitions show opposite orientations, enhancing the discrimination power, as highlighted in Ref.~\cite{Capozzi:2018dat}. Our main goal is to establish a self-calibration pathway for the cross section that is robust against uncertainties in DUNE's final detector performance, the solar flux, and the survival probability.

A precise characterization of the CC $\nu_e+\mathrm{^{40}Ar}$ cross section is important for testing nuclear-physics models, and is a critical prerequisite for unlocking the detector's full potential. Such a measurement would not only allow for a detailed analysis of solar neutrinos~\cite{Capozzi:2018dat}, but would also provide the foundation to unravel supernova explosions~\cite{Mezzacappa:2005ju, Pejcha:2011az, Janka:2012wk, Burrows:2012ew, Adams:2013ana, DUNE:2020zfm, DUNE:2023rtr, DUNE:2024ptd} and probe the low-energy atmospheric neutrino flux~\cite{Battistoni:2005pd, Super-Kamiokande:2021jaq, Zhou:2023mou, Super-Kamiokande:2025sxh}. Above the solar neutrino energy range, inelastic channels involving nucleon emission become non-negligible~\cite{Gardiner:2020ulp}. Consequently, uncertainties increase abruptly, stemming not only from the absence of experimental validation but also from the limitations of current theoretical models. This poses a huge challenge to studies of supernova and atmospheric neutrinos.

The paper is organized as follows. In \Cref{sec:nu40Ar_xsec}, we review the current status of the neutrino-argon cross section. In \Cref{sec:calibration_setup}, we describe the setup we use to perform the cross section calibration in DUNE. In \Cref{sec:Results}, we present the calibration strategies and projected cross section sensitivities. Finally, we draw our conclusions in  \Cref{sec:conclusions}. In the Appendices, we provide additional details and insights.  Throughout, we focus on the solar-neutrino energy range, where the cross section is relatively simple.

\section{Review of the neutrino-argon cross section}\label{sec:nu40Ar_xsec}

In this section, we review the history of developments related to the $\nu_e+\mathrm{^{40}Ar}$ cross section, how it is calculated, its angular distribution, and its de-excitation gamma rays. This interaction is the largest one in liquid argon, though care is needed to separate it from others.  Neutrino-electron scattering (ES) events, $\nu + e^- \rightarrow \nu + e^-$ occur for all flavors of neutrino, though with different weights.  These events can be distinguished statistically by their forward angular distribution and event-by-event if nuclear gamma rays are detected.  There are also neutral-current (NC) neutrino-argon interactions, $\nu + \mathrm{^{40}Ar} \rightarrow \nu + \mathrm{^{40}Ar^*}$, where * indicates an excited state that decays by gamma-ray emission.  These events can be distinguished by their lack of high-energy electrons and potentially by their nuclear gamma rays~\cite{Meighen-Berger:2024xbx}.

\subsection{Brief history of neutrino-argon interactions}
\label{sec:history_nuAr}

Liquid Argon Time-Projection Chamber (LArTPC) detectors were originally proposed in 1977~\cite{Rubbia:1977zz}. This sparked sudden interest in neutrino interactions with argon, leading to the first calculation of MeV-scale neutrino-argon cross section~\cite{Raghavan:1986hv}, which focused on Fermi transitions. This work was immediately followed by evaluations of the capability of LArTPCs to detect solar neutrinos~\cite{Bahcall:1986ry, Raghavan:1986fg}, and later by the first calculation of the GT transitions~\cite{Ormand:1994js}.

The first type of indirect measurements of the transition strengths were performed via $\mathrm{^{40}Ti}$ $\beta^+$ decay~\cite{ICARUS:1998nzl, Bhattacharya:1998hc}. These results were obtained under the assumption of isospin symmetry and negligible differences in the daughter states, inferring the strengths of $\mathrm{^{40}Ar}\rightarrow\mathrm{^{40}K}$ from those of $\mathrm{^{40}Ti}\rightarrow\mathrm{^{40}Sc}$~\cite{Trinder:1997xr}. While the two measurements yielded qualitatively similar results, notable discrepancies remained between them.

One decade later, the second type of indirect measurements of the transition strengths were obtained via $\mathrm{^{40}Ar}(p, n)\mathrm{^{40}K}^*$ scattering~\cite{Bhattacharya:2009zz}, hereafter referred to as $(p,\,n)$. This dataset yields a Gamow-Teller strength distribution that differs substantially from the beta-decay results. The $(p, n)$ results are generally regarded as the most reliable~\cite{Karakoc:2014awa}, and serve as a guide for our analysis. Still, the discrepancy with the beta-decay results introduces additional theoretical uncertainty.

DUNE's scientific program increases the need for theoretical calculations of the cross section. The \texttt{MARLEY} generator~\cite{Gardiner:2020ulp, Gardiner:2021qfr} is often used.  This code combines the above-mentioned indirect measurements at low energies with theoretical approximations at high energies.  In particular, for the energy range needed for a Milky Way supernova, the required calculations must focus on the inelastic regime, in which uncertainties from nuclear physics dominate.  Many papers have addressed this, including Refs.~\cite{Bueno:2003ei, Kolbe:2003ys, GilBotella:2003sz, SajjadAthar:2004yf, SajjadAthar:2005ke, Cheoun:2011zza, Cheoun:2012ha, Tsakstara:2013lca, Paar:2012dj, Dapo:2012pv, Chauhan:2017tgf, Kostensalo:2018kgh, Suzuki:2012ds, Suzuki:2014zea, Pandey:2014tza, VanDessel:2017ery, Gardiner:2018zfg, VanDessel:2019atx, Jachowicz:2019eul, Dolan:2021rdd, Pandey:2025vpa, Nikolakopoulos:2019qcr}. However, there are significant discrepancies among the various approaches in the high-energy regime and inadequate matching to the low-energy regime.  To start to address this, Ref.~\cite{Gardiner:2026psy} makes a promising initial approach to systematize the results, by combining empirical information on low-lying transitions with microscopic calculations of the unbound continuum response, and this is now included in \texttt{MARLEY}. However, these computational advances have not been matched by experimental validation.

As noted above, the DEAP collaboration recently published the first measurement of the neutrino-argon cross section between 12--14.5~MeV~\cite{DEAP:2026img}. 

\subsection{CC $\boldsymbol{\nu_e+\mathrm{^{40}Ar}}$ total cross section}
\label{sec:total_xsec}

The total CC $\nu_e+\mathrm{^{40}Ar}$ cross section at leading order (the allowed approximation) is the sum of independent transitions from $\mathrm{^{40}Ar}$ into different nuclear excited states of $\mathrm{^{40}K}$~\cite{Raghavan:1986hv, Ormand:1994js, Trinder:1997xr, Bhattacharya:1998hc, ICARUS:1998nzl, Bhattacharya:2009zz, Capozzi:2018dat}. Transitions into the ground state of $\mathrm{^{40}K}$ ($Q_\mathrm{gs}=1.504\,\mathrm{MeV}$) are forbidden. The cross section for one of these transitions, $i$, is
\begin{equation} \label{eq:sigma_i}
    \sigma_i(E_\nu) = \frac{G_{F,\beta}^2|V_{ud}|^2}{\pi}  |\mathcal{M}_{o\rightarrow i}|^2 E_e^ip_e^iF(Z,E_e^i)\,,
\end{equation}
where $G_{F,\beta}$ is the Fermi constant for beta decay and $V_{ud}$ is the quark mixing matrix element. The transition-dependent variables are the total electron energy, $E_e^i = E_\nu-Q_\mathrm{gs}-\Delta E_i+m_e$, the total electron momentum, $p_i = \sqrt{(E_e^i)^2-m_e^2}$, the Fermi function accounting for Coulomb effects, $F$~\cite{ Schopper:1969jkp}, and the transition amplitudes~\cite{KonopinskiBook},
\begin{equation}
    |\mathcal{M}_{o\rightarrow i}|^2 = B_i(\mathrm{F})+B_i(\mathrm{GT})\,,
\end{equation}
where the $B_i$ values are indirectly inferred either via $(p,\,n)$ scattering~\cite{Bhattacharya:2009zz} or $\mathrm{^{40}Ti}~\beta^+$ decay~\cite{ICARUS:1998nzl, Bhattacharya:1998hc}. For further details and insights on the calculation, see Ref.~\cite{Capozzi:2018dat}.  As shown in Ref.~\cite{Gardiner:2026psy}, in the solar neutrino energy range, corrections due to nuclear breakup and forbidden transitions are expected to be negligible.

\begin{table}[t]
\renewcommand{\arraystretch}{1.2}
\setlength{\tabcolsep}{5pt}
\centering
\begin{tabular}{||c|c|c|c|c||}
\hline \hline
$i$ & $\Delta E_i$ & $B_i$(F) & $B_i$(GT) & $N_{\text{events}}$ \\
& [MeV] & & & [20\,kton$\cdot$year]  \\
\hline \hline
1  & 2.333 &      & 1.64 & 8038  \\ \hline
2  & 2.775 &      & 1.49 & 5695  \\ \hline
3  & 3.204 &      & 0.06 & 184  \\ \hline
4  & 3.503 &      & 0.16 & 376   \\ \hline
5  & 3.870 &      & 0.44 & 806  \\ \hline
6  & 4.384 & 4.00 &      & 4739  \\ \hline
7  & 4.421 &      & 0.86 & 987   \\ \hline
8  & 4.763 &      & 0.48 & 400  \\ \hline
9  & 5.162 &      & 0.59 & 325  \\ \hline
10 & 5.681 &      & 0.21 & 61  \\ \hline
11 & 6.118 &      & 0.48 & 75 \\ \hline
12 & 6.790 &      & 0.71 & 33 \\ \hline
13 & 7.468 &      & 0.06 & 1    \\ \hline
14 & 7.795 &      & 0.14 & $<1$     \\ \hline
15 & 7.952 &      & 0.97 & 2    \\ \hline \hline
total &    & 4.00 & 8.29 & 21722  \\ \hline \hline
\end{tabular}
\caption{Excitation energies, transition strengths, and expected numbers of events for the relevant transitions of $\nu_e+\mathrm{^{40}Ar}\rightarrow e^-+\mathrm{^{40}K^*}$.  The excitation energy of the superallowed Fermi transition is obtained from Ref.~\cite{Bhattacharya:1998hc}, while its strength corresponds to the theoretical prediction. For the Gamow-Teller excitation energies and strengths we adopt those from Ref.~\cite{Bhattacharya:2009zz}.}
\label{tab:E_and_strengths}
\end{table}

\Cref{tab:E_and_strengths} shows the expected excitation energies, $\Delta E_i$, of each state together with their measured strengths, $B_i$. For GT states, the transitions are those from $(p,\,n)$ measurements~\cite{Bhattacharya:2009zz} with an additional weighting of $g_A^2=(-1.26)^2$ as per convention. For the Fermi transition, the excitation energy is taken from Ref.~\cite{Bhattacharya:1998hc} and the theoretical strength follows from $B_i(F)=N-Z=4$~\cite{Towner:1973yrc, Ormand:1994js, Bhattacharya:2009zz}. The expected number of events in a DUNE-like detector with an exposure of $20\,\mathrm{kton\cdot year}$ is also shown. While the excitation energies are well known, the strength values are not, where a $\lesssim$10\% uncertainty is estimated~\cite{Capozzi:2018dat}.

The total cross section is simply obtained by summing these individual contributions,
\begin{equation}\label{eq:total_xsec_nuAr}
    \sigma_\mathrm{Ar}(E_\nu) = \sum_{i=1}^{N} \sigma_i(E_\nu)\,.
\end{equation}
For a fixed neutrino energy, each transition leads to a discrete electron kinetic energy, $T_e = E_\nu-Q_\mathrm{gs}-\Delta E_i$. For a neutrino spectrum, the electron spectrum is continuous above each threshold.

As discussed in \cref{sec:detection_features}, the ensemble of transitions with close gamma-ray excitation energies together with the continuum neutrino spectrum and the additional smearing of the detector make the electron detected energy spectrum look very similar for all the transitions involved. In addition, the discrimination power worsens due to the lower threshold in detected energy, which is typically chosen as $T_\mathrm{thres}^\mathrm{DUNE}=5$~MeV for DUNE. These low-energy electrons that are missed would have helped disentangle transitions with higher and lower excitation energies.  Below, we briefly comment on the physics prospects if the threshold could be lowered.

\begin{figure}[t]
    \centering    \includegraphics[width=\columnwidth]{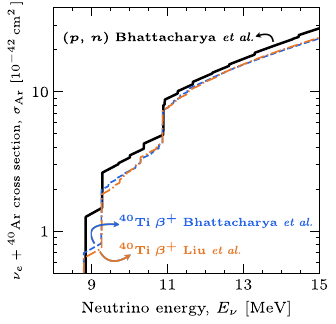}
    \caption{Cross section of the CC $\nu_e+\mathrm{^{40}Ar}$ process, assuming a DUNE threshold of $T_\mathrm{thres}=5\,\mathrm{MeV}$.  \emph{Results from both indirect measurement methodologies generate cross sections that are inconsistent with each other.}}
    \label{fig:cross_section_all_indirect}
\end{figure}

\Cref{fig:cross_section_all_indirect} shows the calculated cross sections using the $(p,\,n)$ values reported in \Cref{tab:E_and_strengths} compared to the two additional $\mathrm{^{40}Ti}~\beta^+$ indirect cases. For the DUNE threshold, the effective neutrino threshold is $E_\mathrm{\nu,\,min}=T_\mathrm{thres}^\mathrm{DUNE}+Q_\mathrm{gs}+\Delta E_{1} \sim9$~MeV for $T_\mathrm{thres}^\mathrm{DUNE} = 5$~MeV. The figure highlights the discrepancies among the different cross sections, which stem from inconsistent transition levels and strengths across the three available indirect measurements. \emph{Although there are arguments in favor of the $(p,\,n)$ measurements~\cite{Karakoc:2014awa}, the ongoing ambiguity and the critical role that LArTPCs will play in the future of neutrino physics underscore the need for a reassessment of the indirect measurements}. Furthermore, establishing a direct measurement to verify the cross section profile represents the definitive path forward, and it is the primary goal of this paper. For a detailed discussion of the various indirect measurement transitions and their inconsistencies, see Ref.~\cite{Gardiner:2020ulp}.

\subsection{CC $\boldsymbol{\nu_e+\mathrm{^{40}Ar}}$ angular distribution}
\label{sec:angular_dist}

The angular distribution of the outgoing electrons is significantly different between Fermi and Gamow-Teller transitions~\cite{Beacom:1998fj, Vogel:1999zy}. Due to the absence of mixed states and the negligible energy dependence of the angular distribution, the cross section's energy and angular dependence can be decoupled and expressed as
\begin{align}\label{eq:xsec_dEdcth}
    \frac{\mathrm{d}^2\sigma_\mathrm{Ar}(E_\nu, \cos\theta)}{\mathrm{d}T_e\mathrm{d}\cos\theta} = \frac{\mathrm{d}\sigma_{\rm F}(E_\nu)}{\mathrm{d}T_e}&\left[\frac{\mathrm{d}\sigma}{\mathrm{d}\cos\theta}\right]_\mathrm{Fermi} \nonumber \\
    +  \sum_{i\in \mathrm{GT}} \frac{\mathrm{d}\sigma_i(E_\nu)}{\mathrm{d}T_e}&\left[\frac{\mathrm{d}\sigma}{\mathrm{d}\cos\theta}\right]_\mathrm{GT}\,,
\end{align}
where the energy distribution of each transition is just
\begin{equation}\label{eq:xsec_Te}
    \frac{\mathrm{d}\sigma_i(E_\nu)}{\mathrm{d}T_e} = \sigma_i(E_\nu) \, \delta\left(T_e-\left[E_\nu-Q_{\rm gs}-\Delta E_i\right]\right)\,,
\end{equation}
with $\delta(x)$ the Dirac delta distribution, and the Fermi angular distribution is
\begin{equation}\label{eq:Fermi_ang}
    \left[\frac{\mathrm{d}\sigma}{\mathrm{d}\cos\theta}\right]_\mathrm{Fermi} = \frac{1}{2} \left( 1 + \cos\theta \right)\,,
\end{equation}
while the Gamow-Teller is
\begin{equation}\label{eq:GT_ang}
    \left[\frac{\mathrm{d}\sigma}{\mathrm{d}\cos\theta}\right]_\mathrm{GT} = \frac{1}{2} \left( 1 - \frac{1}{3}\cos\theta \right)\,.
\end{equation}
These forms are energy-independent because recoil and weak magnetism corrections can be neglected, as they both scale as $\sim$$E_\nu/M$, where $M$ is the nuclear mass. These broad angular distributions are easily within reach of detection in DUNE because the angular resolution is expected to be around 25$\degree$~\cite{Arneodo:2000fa}.  A worse angular resolution would only have modest effects. Below, we provide results for the extreme case of no angular information. In the appendix, we show the effects on the angular distribution upon varying the angular resolution .

Although there is only one Fermi state present, the expected high value of its strength makes it the transition with the third highest expected number of events. This turns out to be crucial when extracting the cross section: although the energy distributions of all transitions are very similar, the Fermi transition can be isolated via its angular distribution.

\begin{figure}[t]
    \centering    \includegraphics[width=\columnwidth]{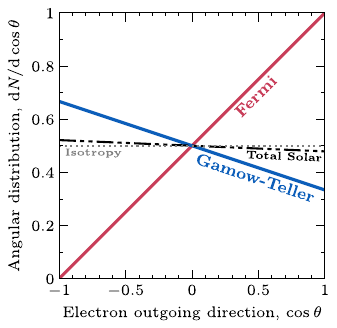}
    \caption{Angular distributions of the outgoing electron for the Fermi and Gamow-Teller transitions, described by \cref{eq:Fermi_ang,eq:GT_ang}, respectively. Although both distributions show distinct orientations with opposite tendency, the weighted combination results in a distribution close to isotropy.}
    \label{fig:angular_dist_F_and_GT}
\end{figure}

\Cref{fig:angular_dist_F_and_GT} shows the angular distributions of the outgoing electrons for $\nu_e+\mathrm{^{40}Ar}$. The Fermi transition shows a forward orientation, while the combination of GT transitions shows a backward orientation. This pattern turns out to be crucial for disentangling both samples. Interestingly, the combination of Fermi and Gamow-Teller events over the relevant energy range leads to a close-to-isotropic sample. The combined result has been obtained with the Gamow-Teller and Fermi normalizations in \Cref{tab:E_and_strengths}~\cite{Bhattacharya:2009zz}, but other cases could be different. \emph{Identifying the Fermi transition would immediately provide the CC $\nu_e+\mathrm{^{40}Ar}$ cross section with directional capability}. This feature could be exploited to both study the angular resolution of the detector with solar neutrinos and for future supernova pointing searches.

\subsection{De-excitation gamma rays}
\label{sec:gammas}

For each transition, de-excitation of $\mathrm{^{40}K}$ leads to the emission of gamma rays. These gamma rays Compton-scatter electrons, which then produce compact, isolated low-energy features known as \emph{blips}, which ArgoNeuT~\cite{ArgoNeuT:2018tvi} and MicroBooNE~\cite{MicroBooNE:2024prh} have been able to reconstruct at MeV energies. These blips should be detectable at DUNE, but the un-detected below-threshold energy depositions might alter the final efficiency of blip detection due to signal smearing~\cite{Castiglioni:2020tsu}.

We consider two scenarios for nuclear gamma-ray detection, given the lack of certainty for DUNE's capabilities. First, a pessimistic one, where they are never detected. To be conservative, this is our default scenario. Second, an optimistic one, where they are perfectly detected and the total energy of the transition can be reconstructed. Since $E_{\gamma,\rm \,dep} = \Delta E_i$, in this scenario one could identify the parent transition for each event. It may be that some gamma-ray activity can be detected for most events but the energy reconstruction is poor. In this case DUNE cannot identify transitions, but the blip distribution would allow for separation from neutrino-electron scattering events. Once DUNE’s blip characterization is established, event samples could be categorized by their associated de-excitation energy. This additional information could enable partial separation of transitions, which can only improve upon our pessimistic assumption. In addition, good blip characterization could enable neutron reconstruction~\cite{Morquecho:2026vta}, which can remove other potential subdominant background sources for our analysis.

\section{Setup for self-calibration}\label{sec:calibration_setup}

In this section, we describe how self-calibration is possible conceptually, detail our assumptions about detection in DUNE, then provide results on the energy spectra and angular distributions of the separate transitions and the total.  For further details, see Appendix \ref{sec:App_Nev}.

\subsection{Why self-calibration is possible}
\label{sec:self_calibration_possible}

Without including detector-specific details, the principal components of the $^8\mathrm{B}$ neutrino event rate in a LArTPC include
\begin{equation}
    \frac{\mathrm{d}^2N}{\mathrm{d}T_e \, \mathrm{d}\cos\theta} \propto \phi_{\mathrm{B}}(E_\nu)  P_{ee}(E_\nu)\frac{\mathrm{d}^2\sigma_\mathrm{Ar}(E_\nu,\,\cos\theta)}{ \mathrm{d}T_e \mathrm{d}\cos\theta}\,,
\end{equation}
where $\phi_{\mathrm{B}}$ represents the solar electron neutrino flux from $^{8}\mathrm{B}$ decay, $P_{ee}$ is the electron survival probability, with matter effects calculated assuming the Standard Solar Model in Ref.~\cite{Acharya:2024lke}, and $\mathrm{d}^2\sigma_\mathrm{Ar}/\mathrm{d}T_e\mathrm{d}\cos\theta$ is the cross section defined in \cref{eq:xsec_dEdcth}, the parameter we aim to extract.

Nominally, these three factors are degenerate with each other. The situation worsens when we include a factor representing the detection (as opposed to interaction) aspects, discussed in the next subsections.  How, then, can we measure the cross section from the event spectrum?

For DUNE, $^8$B solar neutrinos dominate, as the \emph{hep} flux is too small to be observed. The total flux from $^8$B neutrinos has been measured by SNO, and has a current 4\% uncertainty~\cite{SNO:2011hxd}. The spectrum shape has been characterized with $\lesssim$2\% precision~\cite{Winter:2004kf, Longfellow:2023hoj, Acharya:2024lke}, leaving the global uncertainty as a normalization that will be reduced by ES measurements in JUNO or DUNE~\cite{Capozzi:2018dat}. JUNO is already taking data, and by the time DUNE's first far detector modules start operating in 2029~\cite{bishai2025dune}, JUNO expects to have identified $\sim$30,000 signal events~\cite{JUNO:2020hqc}, leading to a sub-percent statistical uncertainty on the $^8$B flux.

In the relevant energy range, the electron neutrino survival probability approaches the matter-dominated asymptotic value $\sim$$\sin^2\theta_{12}$, which is currently known to 2\% precision~\cite{Esteban:2024eli, deSalas:2020pgw, Capozzi:2018dat}, while the solar mass splitting is known at 1\% precision but is not relevant for this analysis. JUNO's extraordinary measurement with only one month of reactor data~\cite{JUNO:2025gmd} gives promising hope to reach the sub-percent precision on both parameters~\cite{JUNO:2022mxj}. For this work we use the global fit of Ref.~\cite{Esteban:2024eli}.

The main expected uncertainty outside the cross section is thus an overall normalization at the sub-percent level. The current uncertainty is at the $\sim$4\% level, meaning that even now we could improve the present 10\% uncertainty of the cross section. Nonetheless, we trust the near-term experimental neutrino program to sufficiently reduce uncertainties, making the assumption of perfect knowledge regarding the flux and mixing parameters a realistic approximation. 

On the detection side, there will be uncertainties, for example in the overall exposure.  Those will increase the total uncertainty on the cross section, but these will need to be treated in separate studies.  For reference, the absolute uncertainty on the solar-neutrino rate achieved by Super-Kamiokande is $\sim$1.5\%~\cite{Super-Kamiokande:2023jbt}.

It will also be important to consider detector backgrounds, following the detailed discussions in Refs.~\cite{Capozzi:2018dat, Zhu:2018rwc, Castiglioni:2020tsu, Parsa:2022mnj, Caratelli:2022llt, Meighen-Berger:2024xbx, Morquecho:2026vta}.  While it is known that some backgrounds can easily be suppressed, for others their impact depends strongly on DUNE's ability to register and separate blips.  First, the captures of external neutrons are a dominant background.  These could be greatly reduced with even modest shielding, which probably has to wait for future DUNE modules. But machine-learning techniques could powerfully reduce this background, as neutron captures produce multiple gamma rays that each scatter multiple electrons, quite different from the $\nu_e+\mathrm{^{40}Ar}$ CC signal, which has a dominant electron track.  Second, there is a class of backgrounds that follow from the ingress of $^{222}$Rn gas; these could be greatly reduced by purification of the air and argon.  Third, there is also a class of backgrounds due to pileup of low-energy events, where again we might expect that machine-learning techniques could be important. Further study is needed.

\subsection{DUNE detection aspects}
\label{sec:DUNE_detector}

We assume two 10-kton liquid argon operational modules for this study, corresponding to $N_{\rm Ar} = 3\times 10^{32}$ argon targets within the fiducial volume~\cite{DUNE:2018tke, DUNE:2018hrq, DUNE:2018mlo, DUNE:2016rla}. Neutrino-argon interactions dominate the signal, but neutrino-electron elastic scattering generates a background that we neglect in this analysis. This contamination may degrade the sensitivity of the analysis, but a modest increase in the observation time would compensate~\cite{Capozzi:2018dat}.

Our main observable for neutrino-argon interaction is the final-state electron kinetic energy~\cite{DUNE:2020zfm, Shi:2025rob}. We assume that DUNE electron detection threshold corresponds to a kinetic energy of $T_e=5~\mathrm{MeV}$~\cite{DUNE:2016rla}, and that electrons could be detected with high efficiency, adopting a constant $\delta T = 7\%$ energy resolution~\cite{DUNE:2024wvj}. We also consider the case of $\delta T= 20\%$. MicroBooNE has demonstrated that energy resolutions within the considered range are achievable in LArTPCs~\cite{MicroBooNE:2023sxs,MicroBooNE:2026kgs}

The primary advantage exploited in this work is the distinct angular electron spectrum of Fermi and Gamow-Teller transitions. We model the angular reconstruction using the von Mises-Fisher distribution~\cite{fisher1995statistical,MardiaJupp,Fisher_disp_sphere}, which provides a rigorous framework for directional data on a sphere. \emph{While a planar approximation remains reasonably accurate, the von Mises-Fisher distribution introduces no additional computational complexity and ensures a correct angular treatment.} A comprehensive derivation of the angular reconstruction formalism is provided in Appendix~\ref{sec:angular_reconstruction}. Based on ICARUS simulations, we adopt an angular resolution of $\theta_{68} = 25\degree$~\cite{Arneodo:2000fa}, which corresponds to a von Mises-Fisher concentration parameter of $\kappa = 12.25$. The final DUNE MeV angular resolution is still unknown, therefore we also evaluate its impact in our analysis by focusing on the limiting case of not including angular information.

\subsection{Detection features}
\label{sec:detection_features}

\begin{figure}[t]
    \centering    
    \includegraphics[width=0.99\columnwidth]{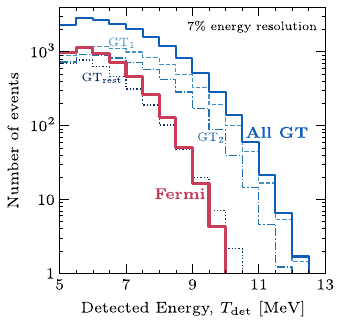}
    \caption{Expected detected electron energy spectra for Fermi and GT transitions for an exposure of 20 kton$\cdot$year. \emph{All states have similar spectrum shapes; those with the lowest thresholds dominate the total spectrum.} States with closer excitation energies exhibit more similar spectrum shapes.}
    \label{fig:energy_spectra}
\end{figure}

\Cref{fig:energy_spectra} highlights the energy distribution of detected electrons for each transition as a function of the detected energy (the electron kinetic energy). The energy spectra are in agreement with those in Ref.~\cite{Capozzi:2018dat}. As shown, all spectra peak at low detected energies and share similar shapes, differing primarily in their high-energy tails. Because the Fermi signal is subdominant to the GT signal, the inclusion of angular information becomes essential for its identification. Another important takeaway from the figure is the hierarchy of GT transitions, where $i=1$ and $i=2$ clearly dominate the remaining contributions due to their lower thresholds.  If it were possible to go to even lower electron energies in DUNE, that could be beneficial for separating the different transitions. However, the highest priority should be on detecting nuclear gamma rays to identify specific transitions, potentially on an event-by-event basis.

\begin{figure}[t]
    \centering    \includegraphics[width=\columnwidth]{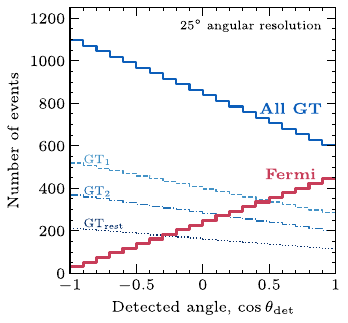}
    \caption{Expected angular distributions of detected electrons relative to the solar direction for an exposure of 20 kton$\cdot$year. \emph{The contrasting angular distributions of the Fermi and GT transitions allow separation of the Fermi signal from the dominant GT signal.}}
    \label{fig:angular_spectra}
\end{figure}

\Cref{fig:angular_spectra} shows the angular distribution of detected events with respect to the direction of the Sun. The Fermi transition has a very distinctive pattern, showing a forward orientation, that allows for clear identification of the Fermi transition under a high-enough exposure. In turn, all GT distributions show the same pattern with different normalizations due to their different excitation energies and transition strengths.

\section{Cross section extraction results}\label{sec:Results}

In this section, we explain our extraction methods and provide the results of our analysis. We examine a variety of scenarios of increasing complexity. We also discuss potential ways to reduce the uncertainty on the cross section via prior knowledge of nuclear transitions and sophisticated reconstruction techniques. All results are obtained assuming the true cross section is characterized by the transition and transition strengths listed in \Cref{tab:E_and_strengths}~\cite{Bhattacharya:2009zz} and for an exposure of 20 kton$\cdot$year.

\subsection{Overall normalization}
\label{sec:overall_norm}

If the spectrum shape of the neutrino–argon cross section were known, the extraction would reduce to a determination of the overall normalization. This measurement would be primarily limited by the total event statistics. With an expected rate of $\sim$20,000 events per 20 kton$\cdot$year exposure, a precise determination could be achieved within a short timeframe. Because the statistical uncertainty scales as $\sim$$1/\sqrt{N}$, \emph{within one year of observation, the neutrino-argon normalization could be determined to below the percent level.} However, this result relies on the combination of 15 independent transitions being in a fixed ratio. Variations in these individual strengths would result in an increase in the uncertainty.

\subsection{Optimistic scenario: perfect identification of nuclear gamma rays}
\label{sec:result_perfectIDs}

If the DUNE far detector allows for high-efficiency detection of the de-excitation gamma ray products of each event, the cross section extraction proceeds in the same manner as the overall normalization but transition-by-transition. In this case as well, the number of expected events is the main factor determining the uncertainty.

\Cref{tab:strength_uncertainties} shows the projected cross section transition strength uncertainties under this optimistic scenario. The first nine transitions are very well constrained, while there is very poor knowledge of the last ones. Indeed, following the discussion in Refs.~\cite{Capozzi:2018dat,Bhattacharya:2009zz}, one could safely assume that the transition strength has a $10\%$ uncertainty. This highlights the importance of a new assessment on indirect measurements of neutrino-argon interactions, since it will be crucial in order to achieve the needed precision level for future physics searches. Relying on these assumptions, \emph{a 1\% determination of the cross section is achievable provided that DUNE can detect de-excitation signatures with high efficiency.} In addition, the transitions with $i= 1$, 2, and 6 (Fermi) will be determined at the $\lesssim 2\%$ level, allowing direct access to the dominant GT and the Fermi transition strengths.

\begin{table}[t]
\renewcommand{\arraystretch}{1.2}
\setlength{\tabcolsep}{5pt}
\centering
\begin{tabular}{||c|c|c|c|c||}
\hline \hline
$i$ & $B_i(F)$ & $B_i(GT)$ & \multicolumn{2}{c||}{$\delta B_{i}^{\mathrm{opt}}$ [\%]} \\ 
 & & & \multicolumn{1}{c}{($\delta T = 7\%$)} & \multicolumn{1}{c||}{($\delta T = 20\%$)} \\ \hline \hline
1 & & 1.64 & 1.1 & 1.2 \\ \hline
2 & & 1.49 & 1.3 & 1.4 \\ \hline
3 & & 0.06 & 7.5 & 8.1 \\ \hline
4 & & 0.16 & 5.2 & 5.6 \\ \hline
5 & & 0.44 & 3.6 & 3.9 \\ \hline
6 & 4.00 & & 1.5 & 1.6 \\ \hline
7 & & 0.86 & 3.2 & 3.5 \\ \hline
8 & & 0.48 & 5.1 & 5.6 \\ \hline
9 & & 0.59 & 5.7 & 6.2 \\ \hline
10 & & 0.21 & 13 $(10^*)$ & 15 $(10^*)$ \\ \hline
11 & & 0.48 & 12 $(10^*)$ & 13 $(10^*)$ \\ \hline
12 & & 0.71 & 18 $(10^*)$ & 21 $(10^*)$ \\ \hline
13 & & 0.06 & 210 $(10^*)$ & 250 $(10^*)$ \\ \hline
14 & & 0.14 & 240 $(10^*)$ & 280 $(10^*)$ \\ \hline
15 & & 0.97 & 100 $(10^*)$ & 120 $(10^*)$ \\ \hline \hline
\end{tabular}
\caption{Relative uncertainties for each transition strength under the optimistic scenario. The relative uncertainty for all transitions is capped at the 10\% uncertainty derived from indirect measurements.}
\label{tab:strength_uncertainties}
\end{table}

\subsection{Pessimistic scenario: no detection of nuclear gamma rays}
\label{sec:result_noGammaRays}

Now,  we follow an intuitive approach adopted to determine the cross section without de-excitation gamma-ray information.

Our extraction approach groups the 15 transitions into three effective groups, informed by the strengths and energies in \Cref{tab:E_and_strengths} and the experimental features of the transition strengths, as shown in \Cref{fig:angular_dist_F_and_GT}. This reduction results in: ($i$) a combined set of the first two transitions, which provides the highest statistics but remains highly degenerate due to their close excitation energies and strengths, ($ii$) the Fermi transition, distinguished by its unique angular distribution, and ($iii$) the remaining Gamow-Teller transitions, which collectively contribute a total event count comparable to that of the Fermi transition. This is,
\begin{equation}\label{eq:3eff}
    \sigma_\mathrm{Ar} = n_{1+2}(\sigma_1+\sigma_2) + n_{\rm F}\sigma_6 + n_\mathrm{rest}\sum_{i>2,\,i\neq6}^{15}\sigma_i\,,
 \end{equation}
where $n_i$ are normalizations to the combinations of $\sigma_i$ with \cref{tab:E_and_strengths} strengths, as defined in \cref{eq:sigma_i}. Although defined here as a three-component extraction approach, we remark that this methodology does not utilize only three transitions, it aggregates the full set of indirectly measured transitions into three functionally distinct groups.

Under this approach, one can forecast the sensitivity to the normalizations in \cref{eq:3eff} assuming that the true scenario is that of the indirect strengths determination. In such scenario, $n_i=1$, and the relative uncertainties, $\delta n_i \equiv \sigma_{n_i}/n_i$, where $\sigma_{n_i}$ is the absolute uncertainty, are
\begin{align}
    \delta n_{1+2} &=  2.0\,\%\,, \nonumber\\
    \delta n_{\rm F} &= 4.2\,\% \,, \nonumber \\
    \delta n_\mathrm{rest} &= 9.9\,\% \,. \nonumber 
\end{align}
There is no degeneracy between $n_{1+2}$ and $n_{\rm F}$ due to their high statistics, distinct excitation energies, and contrasting angular distributions, allowing for an independent extraction. However, both components exhibit a slight degeneracy with $n_\mathrm{rest}$. Specifically, $n_{1+2}$ shares the same angular distribution as $n_\mathrm{rest}$ and lacks sufficient spectrum separation in energy, while $n_{\rm F}$ possesses an excitation energy similar to the effective energy of the $n_\mathrm{rest}$ cluster, resulting in a mild correlation despite their very different angular spectra.

\begin{figure}[t]
    \centering    \includegraphics[width=\columnwidth]{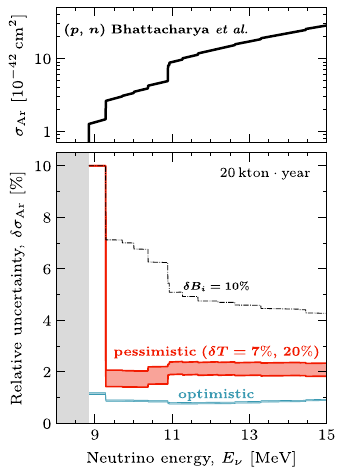}
    \caption{Cross section of the CC $\nu_e+\mathrm{^{40}Ar}$ process and its projected relative uncertainty. The bottom panel shows the relative uncertainty for the optimistic (perfect identification) and pessimistic (no gamma-ray detection) scenarios. The lower and upper limits for each are marked by 7\% and 20\% energy resolution, respectively. The optimistic result incorporates a 10\% cap on individual strength uncertainties. \emph{DUNE will be able to constrain the cross section at the few percent level using solar neutrinos}.}
    \label{fig:nuAr_xsec_result}
\end{figure}

\Cref{fig:nuAr_xsec_result} illustrates the uncertainty in the cross section derived from the three-component transition strengths analysis in \cref{eq:3eff}. \emph{Although this approach assumes the most pessimistic DUNE setup, a cross section determination with a precision of $\lesssim$2\% remains achievable across nearly the entire energy range}. The primary advantage of this approach is the grouping of the dominant transitions, $\sigma_1$ and $\sigma_2$, into a single component to account for their strong degeneracy. This allows for a robust determination of the overall cross section normalization by constraining the linear combination $\sigma_1+\sigma_2$, to which the analysis is highly sensitive, without requiring independent knowledge of each individual strength. The unique electron angular signature of the Fermi transition enables its clear separation from the GT signal. In addition, we account for the minor contributions of the remaining Gamow-Teller transitions by grouping them into a single component. Despite the higher uncertainties associated with these transitions, the constraints placed on the dominant transitions effectively limit the total cross section uncertainty to $\sim$2\%.

The effect of energy resolution is also visible, where the upper bound of the pessimistic scenario is obtained with 20\% energy resolution for electrons, while the lower limit is the projected sensitivity assuming 7\%. Without accounting for angular information, the cross section extraction in the three effective transitions approach is impossible to achieve.

We tested the cross section reconstruction performance by generating random transition strengths within 10\% of the $(p,n)$ benchmark. In analyzing the subsequent mock data, the underlying cross section was always recovered within the uncertainty. This further validates our calibration approach.

\subsection{Towards more precision: PCA}
\label{sec:more_precision_PCA}

In the following, we make use of Principal Component Analysis (PCA) to identify data-driven signal combinations and perform the cross section extraction over the most relevant combinations of transitions. Finally, a hybrid approach is implemented, which combines the direct extraction of the Fermi transition strength with a PCA-based treatment of the remaining components. PCA provides a rigorous framework to guide our analysis, maximizing our extraction capabilities.

\subsubsection{Principal Component Analysis method}
\label{sec:result_2PCA}

PCA not only identifies the principal components of the cross section sensitivity but also quantifies the number of optimal degrees of freedom in a purely data-driven manner. By incorporating both energy and angular information, we find that the effective dimensionality of the data is two. This indicates that the three effective transitions approach assumes a higher dimensionality than the data can support, suggesting that a more optimized, low-dimensional analysis could further improve our results without risking an accurate reconstruction of the cross section.

By reducing the whole dimensionality of the extraction analysis to two, the PCA method suggests that our cross section optimal configuration has the form
\begin{equation}
    \sigma_\mathrm{Ar} = n_1(\vec{v}_1\cdot\vec{\sigma}) + n_2(\vec{v}_2\cdot\vec{\sigma})\,,
\end{equation}
where $\vec{\sigma}$ is a vector with the 15 different transitions as elements, and $\vec{v}_i$ is a normalized vector containing weights for each of the cross section transitions. The PCA method also provides the specific vectors containing the weights of each cross section transition, $\vec{v}_i$. For more details on the PCA approach and its application to the cross section extraction, see Appendix~\ref{sec:App_PCA}.

Specifically, as the cross section is expected to align closely with the indirect measurements in Ref.~\cite{Bhattacharya:2009zz}, we parameterize the cross section in terms of its deviation from the expected central value:
\begin{equation}
    \sigma_\mathrm{Ar} = (\vec{\mathbf{1}} + n_\mathrm{1,pca}\vec{v}_\mathrm{1,pca} + n_\mathrm{2,pca}\vec{v}_\mathrm{2,pca})\cdot\vec{\sigma}\,,
\end{equation}
where $\vec{\mathbf{1}}$ is an identity vector of the same dimension as the transition vector $\vec{\sigma}$. By adopting this framework, the best-fit values for the coefficients $n_\mathrm{1,pca}$ and $n_\mathrm{2,pca}$ are expected to be zero if the results perfectly match the prior experimental data. In this extraction scenario, the weighting vectors are derived entirely via the PCA method, ensuring an optimized, data-driven parameterization of the cross section.

The projected sensitivity for the coefficients $n_\mathrm{1,pca}$ and $n_\mathrm{2,pca}$ assuming an exposure of 20 kton$\cdot$year is
\begin{align}
    \delta n_{1,\mathrm{pca}} &=  1.4\,\%\,, \nonumber\\
    \delta n_{2,\mathrm{pca}} &= 3.7\,\% \,. \nonumber 
\end{align}

\begin{figure}[t]
    \centering    \includegraphics[width=\columnwidth]{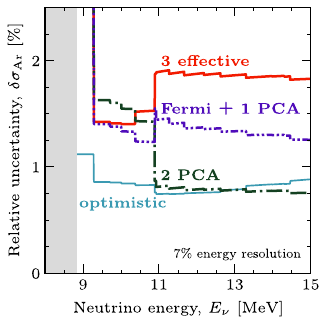}
    \caption{Projected relative uncertainties of the cross section for the various reconstruction methods discussed in this work, for a 20 kton$\cdot$year exposure. The different methodologies yield consistent results, with uncertainties remaining below 2\%, excluding the first transition. \emph{The limited difference between the pessimistic and optimistic scenarios suggests that the cross section reconstruction remains highly feasible, regardless of the final specifications of DUNE's experimental configuration}.}
    \label{fig:all_results_ratio}
\end{figure}

\Cref{fig:all_results_ratio} presents the projected cross section uncertainty derived from the PCA method. We observe significant precision gains when transitioning from the three effective transitions grouping to the 2 PCA approach, which reduces the uncertainty by approximately a factor of $\sim$2 at the 11--15~MeV range. The primary advantage of this method is its purely data-driven nature. However, while this approach is statistically optimal for cross section extraction, the resulting principal components lack direct physical intuition. Nevertheless, one could still map each individual strength value and its uncertainty from the combination of the two PCA vectors. \emph{By maximizing the information gain through the PCA method, we achieve a high-precision measurement of the cross section, with uncertainties of $\lesssim$2\% below the Fermi transition energy and $\lesssim$1\% at higher energies.} To account for the high degree of degeneracy between states $i=1$ and $i=2$, the associated cross section uncertainty corresponding to the first transition (first step) is kept at the 10\% level, following the indirect measurements result. This ensures a robust assessment of the reconstruction performance and prevents over-optimistic reconstruction results.

\subsubsection{Fermi strength + 1 PCA}
\label{sec:result_Fermi+PCA}
While PCA might be the optimal method to extract the cross section, the lack of physical intuition and sensitivity to individual strengths makes the PCA approach less appealing if there is not a substantial gain of sensitivity. Here, our primary objective is to extract the $\nu_e+\mathrm{^{40}Ar}$ CC cross section while maintaining physical interpretability. To achieve this, we move away from a purely data-driven model and partition the data into two subsets based on their most distinct physical signature: the electron angular distribution. This allows us to isolate a Fermi-dominated sample and a GT sample. To preserve high constraining power within the GT sector, we apply PCA to the GT sample, identifying the dominant direction of variance. This leads to a cross section
\begin{equation}
    \sigma_\mathrm{Ar} = ( 1+ n_{\rm F} )\sigma_{\rm F} + (\vec{\mathbf{1}} +n_{\rm GT}\vec{v}_{\rm GT})\cdot\vec{\sigma}_{\rm GT}\,,
\end{equation}
where $\vec{\sigma}_{\rm GT}$ is the subset of $\vec{\sigma}$ excluding the Fermi component, and $\vec{v}_{\rm GT}$ represents the principal component vector that captures the dominant direction of variance within the Gamow-Teller data. This approach preserves the two-dimensional fashion while providing a clear physical understanding of each vector.

The projected sensitivity of DUNE for the two parameters in the Fermi + 1 PCA approach are
\begin{align}
    \delta n_{\rm GT} &=  2.0\,\%\,, \nonumber\\
    \delta n_{\rm F} &= 3.5\,\% \,. \nonumber 
\end{align}
While the sensitivity levels are comparable to those achieved in the 2 PCA approach, these parameters have a clear physical meaning. One of these parameters is directly the normalization of the Fermi strength, gaining direct access to one of the transitions, while the other one is a normalization of the weighted GT strengths.

\Cref{fig:all_results_ratio} presents the projected cross section sensitivity for this hybrid approach, facilitating a direct comparison with the other methodologies. This case achieves an uncertainty reduction of approximately a factor of 1.5 across the entire energy spectrum compared to the three effective transitions grouping. Relative to the 2 PCA scenario, the hybrid approach offers superior precision at lower energies but suffers an increased uncertainty at energies above the Fermi transition. \emph{While the Fermi transition strength is resolved to within 3.5\%, this comes at the cost of higher cross section uncertainty at higher energies.} In addition to isolating the Fermi strength, this method effectively encapsulates the uncertainties associated with the axial-vector coupling in the Gamow-Teller sector. This partition facilitates testing for deviations from the nominal value of $g_A=-1.26$. Such variations would manifest as a normalization shift captured by the $n_{\rm GT}$ parameter. Considering the combined advantages of high precision and interpretability, we favor the Fermi + 1 PCA extraction approach over the 2 PCA. Nevertheless, using all three approaches described in this paper provides deeper insight, as each offers unique advantages for characterizing the $\nu_e+\mathrm{^{40}Ar}$ cross section.

\subsubsection{PCA power and additional possibilities}
\label{sec:PCAresult_additional}

A further advantage of the PCA approach over the three effective transitions model is its superior stability against degraded energy resolution. Uncertainties remain below 2\% even under highly pessimistic resolution assumptions, effectively establishing this value as an upper bound for the sensitivity projection. 

\begin{figure}[t]
    \centering    \includegraphics[width=\columnwidth]{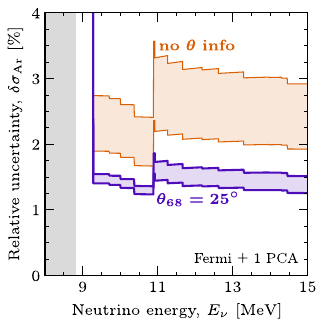}
    \caption{Impact of angular information and energy resolution for the Fermi + 1 PCA case, with 20 kton$\cdot$year exposure and $\delta T=7\%,20\%$. The PCA method enables a good extraction of the cross section without angular information. Including angular information, we project a sensitivity below 2\%, even under a pessimistic $\delta T=20\%$. \emph{By guiding our analysis with PCA, our sensitivity remains at the few percent level, even for pessimistic detector performance.}}
    \label{fig:energy_resolution}
\end{figure}

\Cref{fig:energy_resolution} 
illustrates the constraining power of the PCA method by comparing the sensitivity under 7\% and 20\% energy resolution to a scenario where angular information is omitted. While all results remain below 4\%, the inclusion of angular information substantially enhances the sensitivity due to a greater capacity to resolve the Fermi transition strength. The result without angular information approximates the limiting case of extremely bad angular resolution; this is approximate because one would also need to take into account moderate changes to the assumed backgrounds due to ES events.  Any changes in our results due to a realistically different angular resolution would be small.

While we focus on the three aforementioned approaches, alternative groupings of weighted transitions are possible and could yield comparable results. One could isolate the first and second transitions and parameterize the remaining states using PCA. Given the complexity of the initial 15-state problem compared to the two independent degrees of freedom supported by the data, one has the flexibility to define a specific combination of states, leaving the PCA to identify the optimal parameter for the remainder. However, caution is required; choosing a vector that lacks a strong physical or data-driven basis may lead to overestimating uncertainties, particularly if the chosen vector is poorly constrained by the available data.

In summary, by utilizing PCA to optimize our constraining power, we achieve higher precision on the cross section extraction. We suggest that future studies consider alternative weighting schemes if specific physical motivations arise, as the flexibility of this framework allows for further optimization.

\subsection{Intermediate scenario: partial detection of nuclear gamma rays}
\label{sec:result_some_gammas}

Once DUNE’s blip characterization is known, future analyses could implement Monte Carlo methods to introduce observed gamma-ray energy as a discriminating variable. Constructing a binning scheme including the gamma-ray energy would likely improve precision. A conservative approach would bin the data according to the known excitation levels, allowing for the removal of transitions with lower excitation energies. \emph{The strong degeneracy between the first and second transitions could be reduced by incorporating gamma-ray information.}

A PCA-guided analysis incorporating gamma-ray information could increase the number of independent parameters to which the data is sensitive. Any additional information will only reduce the uncertainty compared to the pessimistic result.

Following the prospects in Ref.~\cite{Castiglioni:2020tsu}, a 10\% total de-excitation gamma-ray energy resolution may be possible. This would provide an excellent capability for resolving neighbor transitions. If so, the projected sensitivity would approach the optimistic scenario.

\section{Conclusions and future work}\label{sec:conclusions}

With DUNE on the horizon, very exciting opportunities arise from having a detector whose main detection target is electron neutrinos. However, successfully developing its MeV neutrino program requires precise knowledge of the dominant detection processes. Despite being DUNE’s primary interaction channel at MeV energies, the CC $\nu_e+\mathrm{^{40}Ar}$ cross section has yet to be precisely constrained. While indirect measurements exist for the individual transitions that constitute the total cross section~\cite{ICARUS:1998nzl, Bhattacharya:1998hc, Bhattacharya:2009zz}, significant experimental and theoretical uncertainty persists. Resolving these discrepancies is essential, as interpreting data from a future galactic supernova explosion~\cite{Mezzacappa:2005ju, Pejcha:2011az, Janka:2012wk, Burrows:2012ew, Adams:2013ana, DUNE:2020zfm, DUNE:2023rtr, Newmark:2023vup, DUNE:2024ptd} or characterizing the low-energy atmospheric neutrino flux~\cite{Battistoni:2005pd, Super-Kamiokande:2021jaq, Zhou:2023mou, Super-Kamiokande:2025sxh} relies entirely on an accurate understanding of this cross section at tens of MeV~\cite{DUNE:2023rtr}.  For example, DUNE could provide a direct measurement of the neutronization peak~\cite{Kachelriess:2004ds, DUNE:2020zfm}, unlocking unprecedented insight into the complex dynamics of supernova explosions, but only if the cross section is known. Our work is an important step towards this goal.

In this paper, we provide DUNE the ability to self-calibrate the CC $\nu_e+\mathrm{^{40}Ar}$ cross section using $^8\mathrm{B}$ solar neutrinos, while waiting for a future stopped-pion experiment to directly address this issue~\cite{Capozzi:2018dat}. We do it for optimistic and pessimistic scenarios on the detection of nuclear gamma rays. Our main goal is to extract the cross section while maximizing information regarding transition strengths. For an exposure of 20 kton$\cdot$year, \emph{our findings demonstrate that a 2\% determination of the cross section within the 9--15~MeV energy range is achievable, even in the most pessimistic experimental scenario.} Detection uncertainties will increase this, but those should be the subject of future studies.

Several strategic directions could follow from our work. First, establishing precise cross-section data up to 15 MeV is highly valuable. It provides a reliable low-energy baseline to check nuclear models and to allow future experiments to directly extract information about the higher-energy regime, where the impact of nuclear breakup processes and forbidden transitions becomes important~\cite{Gardiner:2026psy}.  Second, a deeper theoretical and experimental understanding of higher-energy transitions could extend the utility of this calibration to a broader neutrino energy range, potentially enabling it at higher energies using alternative neutrino sources. Finally, if the $\nu_e+\mathrm{^{40}Ar}$ is precisely characterized, the $\bar{\nu}_e+\mathrm{^{40}Ar}$ CC cross section could be measured during a galactic supernova burst or with atmospheric neutrinos~\cite{Bueno:2003ei, GilBotella:2003sz}.  We leave this exploration for future work, as the first excitation energy for the antineutrino channel is high.


\bigskip
\textbf{\textit{Note added:}} As this paper was being completed, we learned of an independent study by Cheng, Hostert, Machado, Mishra, and Thompson~\cite{SNcandles2026}.  Both contributions were simultaneously submitted to arXiv.  Our two studies have connected but complementary approaches and results.


\begin{acknowledgments}

We are grateful for helpful discussions with Alejandro Garcia, Inés Gil Botella, Scott Haselschwardt, Bryce Littlejohn, Nityasa Mishra, and Vishvas Pandey.

The work of O.N. and J.F.B. was supported by National Science Foundation Grant No.\ PHY-2310018.

J.F.B. speaks for himself as a theorist, not on behalf of the DUNE Collaboration. This work is based on the ideas and calculations of the authors, plus publicly available information.

\end{acknowledgments}


\clearpage
\onecolumngrid
\appendix
\renewcommand\thefigure{\thesection\arabic{figure}}

\setcounter{figure}{0}

{\centering\huge\textsc{Appendices}\par}
\vspace{2em}

In the following, we describe in detail the most relevant aspects of our analysis. In Appendix \ref{sec:App_Nev}, we provide the energy, angular and statistical treatment. In Appendix \ref{sec:App_12deg}, we comment on the most limiting factor of our pessimistic analysis: the degeneracy between the first two Gamow-Teller states. Finally, in Appendix \ref{sec:App_PCA}, we describe the PCA method used in the analysis in more detail.

\vspace{3em}

\section{Analysis pipeline}\label{sec:App_Nev}

\subsection{Basic setup}
\label{sec:basic_setup}

In this study, we focus exclusively on the $\nu_e+\mathrm{^{40}Ar}$ charged-current interaction in DUNE, employing the kinetic energy and angular distribution of the outgoing electron to perform the cross section extraction. 
The number of events within a particular electron kinetic energy $[T_{i}^\mathrm{low},T_{i}^\mathrm{up}]$ and angular range $[\cos\theta_{i}^\mathrm{low},\cos\theta_{i}^\mathrm{up}]$ is obtained as
\begin{equation}
\label{eq:event_rate}
N_{\mathrm{ev},\,i} = t_\mathrm{obs}\times N_\mathrm{Ar}\times\left[\int_{T_{i}^\mathrm{low}}^{T_{i}^\mathrm{up}}\frac{\mathrm{d}R}{\mathrm{d}T_{\rm det}}\mathrm{d}T_{\rm det}\right] \times \left[\int_{\cos\theta_{i}^\mathrm{low}}^{\cos\theta_{i}^\mathrm{up}}\frac{\mathrm{d}R}{\mathrm{d}\cos\theta_{\rm det}} \mathrm{d}\cos\theta_{\rm det} \right]\,,
\end{equation}
where we always work assuming a 20~kton$\cdot$year exposure, corresponding to $t_\mathrm{obs}=1\,\mathrm{year}$ and $N_{\rm Ar} = 3\times 10^{32}$ targets.

\begin{figure}[b]
    \centering    \includegraphics[width=\columnwidth]{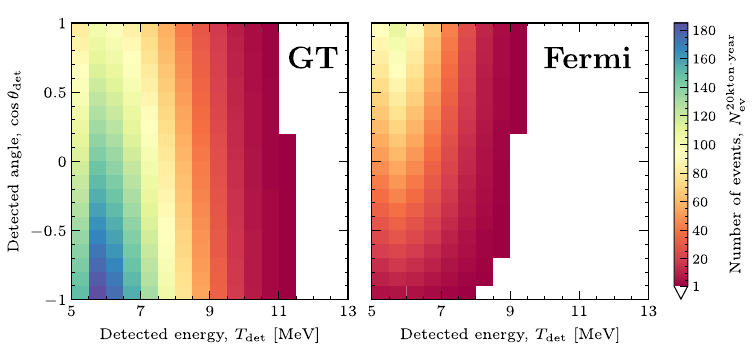}
    \caption{Projected detected angular and energy distributions of all the GT transitions and the Fermi one.}
    \label{fig:Nev_GT_and_F}
\end{figure}

\Cref{fig:Nev_GT_and_F} shows the expected Gamow-Teller and Fermi events for an exposure of 20 kt$\cdot$year. By performing the integrals over the reconstructed energy and angle in \cref{eq:event_rate}, one finds the final event count per bin. We obtain this quantity transition by transition, and later combine them depending on our assumed gamma-ray tagging capabilities. In our analysis, we use bins of $\Delta T_{\rm det}=0.5$~MeV and $\Delta \cos\theta_{\rm det}=0.1$.

In the following, we describe in detail the computation of the angular and energy event distributions, using DUNE's features described in \cref{sec:DUNE_detector}.

\subsection{Detected energy}
\label{sec:energy_reconstruction}

The first term in brackets in \cref{eq:event_rate} contains the energy-dependent rate. This is the most complex term, as it accounts for the entire neutrino journey from emission to detection. It encodes information regarding the emission, transport and detection processes. However, given the precise expected characterization of solar neutrino emission from  
$^8\mathrm{B}$ decay and of the solar mixing parameters, this term provides information on the $\nu_e+\mathrm{^{40}Ar}$ detection process.

The observable energy in DUNE is the electron kinetic energy, $T_\mathrm{true} = E_e - m_e$, which for a specific transition $i$ corresponds to $(T_\mathrm{true})_i = E_\nu - Q_\mathrm{gs} - \Delta E_i$. In terms of these quantities, the differential event spectrum as a function of its reconstructed energy is given by
\begin{equation}
\label{eq:eventrateE}
\frac{\mathrm{d}R}{\mathrm{d}T_{\rm det}} = \int \mathrm{d}T_\mathrm{true} \mathrm{d}E_\nu \mathcal{R}_E\left(T_{\rm true}, T_{\rm det}\right)  \frac{\mathrm{d}\Phi^{\rm D}_{\nu}(E_\nu)}{\mathrm{d}E_\nu} \frac{\mathrm{d}\sigma(E_\nu, T_{\rm true})}{\mathrm{d} T_{\rm true}} \,,
\end{equation}
where ${d\Phi^{\rm D}_{\nu}}/{dE_\nu}$ is the neutrino flux reaching the detector, ${d\sigma}/{d T_{\rm true}}$ is the cross section in \cref{eq:xsec_Te}, and the energy resolution function is assumed to be Gaussian,
\begin{equation}
\mathcal{R}_E\left(T_{\rm true}, T_{\rm rec}\right) = \, \frac{1}{\sqrt{2\pi} \delta T}\exp\left(-\frac{\left(T_{\rm true}-T_{\rm det}\right)^2}{2 \, \delta T^2}\right)\,,
\label{eq:gauss}
\end{equation}
where we take a constant energy resolution of the electron kinetic energy of either $\delta T=7\%$ or $\delta T=20\%$~\cite{DUNE:2016rla}.

Throughout this work, we use an energy binning of $\Delta T_\mathrm{det}=0.5\,\mathrm{MeV}$, and we consider a reconstructed energy interval between 5 and 18 MeV. The lower limit corresponds to DUNE's lower detection threshold, while the upper limit is slightly higher in energy than the $^{8}\mathrm{B}$ endpoint.

\subsection{Angular smearing}
\label{sec:angular_reconstruction}

The last term in \cref{eq:event_rate} contains the angular distribution of the observed event rate. In this case, it corresponds to the detected angular distribution of the electrons, and it reads as~\cite{Costantini:2004ry, Tomas:2003xn}
\begin{equation}
\label{eq:angular_part_Nev}
\frac{\mathrm{d}R}{ \mathrm{d}\cos\theta_{\rm det}} = \, \frac{1}{4\pi} \int \mathrm{d}\cos\theta_\mathrm{true} \mathrm{d}\phi \mathcal{R}_\theta(\theta_\mathrm{true},\theta_\mathrm{det}) \frac{\mathrm{d}\sigma}{\mathrm{d}\cos\theta_\mathrm{true}}\,,
\end{equation}
where $\theta_\mathrm{true}$ and $\phi$ are the true zenith and relative azimuth angles of the emitted electron, $\mathrm{d}\sigma/\mathrm{d}\cos\theta_\mathrm{true}$ is the angular distribution of the electrons, and $\mathcal{R}_\theta$ is the angular resolution function. For angular data in a sphere, the exact angular resolution function is given by the 
probability density of the von Mises-Fisher (vMF) distribution~\cite{fisher1995statistical, MardiaJupp, Fisher_disp_sphere}
\begin{equation}
    \mathcal{R}_\mathrm{vMF}(\gamma; \kappa) = \frac{\kappa}{4\pi\sinh\kappa} \exp\left(\kappa\cos\gamma\right)\,,
\end{equation}
where $\gamma$ represents the angular separation between the true and detected directions, $\cos\gamma = \cos\theta_\mathrm{true} \cos\theta_\mathrm{det} + \sin\theta_\mathrm{true} \sin\theta_\mathrm{det} \cos\phi$, and $\kappa$ is the concentration parameter, which quantifies the degree of clustering around the mean direction.

Since the azimuth is not a variable that contains relevant information in this analysis, one could integrate it out,
\begin{equation}
    f_\mathrm{vMF}(\theta_\mathrm{rec};\theta_\mathrm{true},\kappa) = \frac{\kappa}{2\sinh\kappa} I_0(\kappa\sin\theta_\mathrm{rec}\sin\theta_\mathrm{true})\exp\left(\kappa\cos\theta_\mathrm{rec}\cos\theta_\mathrm{true}\right)\sin\theta_\mathrm{rec}\,,
\end{equation}
where $I_0$ is a modified Bessel function of the first kind. This simplifies \cref{eq:angular_part_Nev} into
\begin{equation}
\frac{\mathrm{d}R}{ \mathrm{d}\cos\theta_{\rm det}} = \, \frac{1}{4\pi} \int \mathrm{d}\cos\theta_\mathrm{true} f_\mathrm{vMF}(\theta_\mathrm{det};\theta_\mathrm{true},\kappa) \frac{\mathrm{d}\sigma}{\mathrm{d}\cos\theta_\mathrm{true}}\,.
\end{equation}

The next step in this process is giving a value to the concentration parameter. We are aware that the numbers for this parameter do not have an easy intuition, but the main trend is that a higher concentration parameter indicates a distribution that is more concentrated towards the mean direction. Since in physics we are used to Gaussian variables, one could find the equivalence between the concentration parameter corresponding to the angular resolution describing 68\% of events contained in a cone of certain opening, which in our benchmark case is taken as $\theta_{68}=25\degree$~\cite{Capozzi:2018dat}. 

In order to do so, one should compute the integral of the vMF distribution and compute the value of $\kappa$ that makes
\begin{equation}
    \mathrm{CDF}_\mathrm{vMF}(\theta_{68}, \kappa) = 0.68\,,
\end{equation}
where the integral of the vMF distribution function is the Cumulative Distribution Function (CDF), that reads as
\begin{equation}
    \mathrm{CDF}_\mathrm{vMF}(\gamma, \kappa) = \frac{1-\exp(\kappa(\cos\gamma-1))}{1-\exp(-2\kappa)}\,.
\end{equation}
%

\begin{figure}[t]
    \centering    \includegraphics[width=0.4\columnwidth]{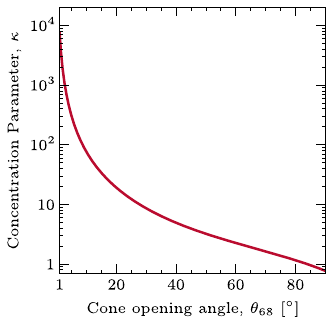}
    \caption{Concentration parameter as a function of the opening angle of the cone that contains 68\% of the events. This relation translates our insight from Gaussian variables to the actual parameter describing the distribution for spherical data.}
    \label{fig:concentration_param}
\end{figure}

\Cref{fig:concentration_param} illustrates the concentration parameter as a function of certain angular resolutions in degrees.
For opening angles $\theta_{68}\lesssim 40\degree$, this relation could be approximated as
\begin{equation}
    \kappa \simeq k\,\theta_{68}^{-2} \,,
\end{equation}
where for $\theta_{68}$ in degrees, $k=7535$. This equation yields values of $\kappa$ that are within $5\%$ of the true value as long as $\theta_{68}\lesssim40\degree$, and has the expected dependence on $\theta_{68}$, $\kappa\propto \theta_{68}^{-2}$~\cite{MardiaJupp}.

In this particular analysis, we focus on the Fermi and Gamow-Teller angular distributions, described in \cref{eq:Fermi_ang,eq:GT_ang}. The angular resolution is assumed to be $\theta_{68}=25\degree$, corresponding to a concentration parameter of $\kappa=12.5$. 

\begin{figure}[b]
    \centering    \includegraphics[width=0.7\columnwidth]{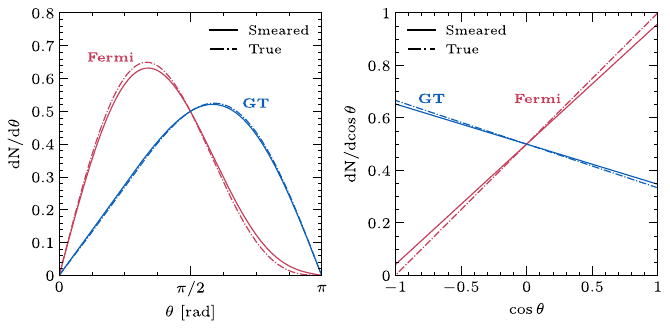}
    \caption{Effect of smearing to the Fermi and Gamow-Teller angular distributions. The smearing is performed using the von Mises-Fisher distribution.}
    \label{fig:AngularSmearing}
\end{figure}

\Cref{fig:AngularSmearing} shows the final reconstructed angular distributions once the angular smearing of the detectors is taken into account. We use $\theta_{65}=25\degree$, but the final angular resolution is uncertain and might lead to small changes in the outcome of the analysis.  

In fact, we recommend abandoning the planar approximation for directional information in favor of the correct angular treatment. The exact approach is no more computationally expensive, which is critical since the planar approximation introduces significant deviations at angular resolutions comparable to those expected for DUNE based on the ICARUS analysis~\cite{Arneodo:2000fa}. 

\begin{figure}[t]
    \centering    \includegraphics[width=\columnwidth]{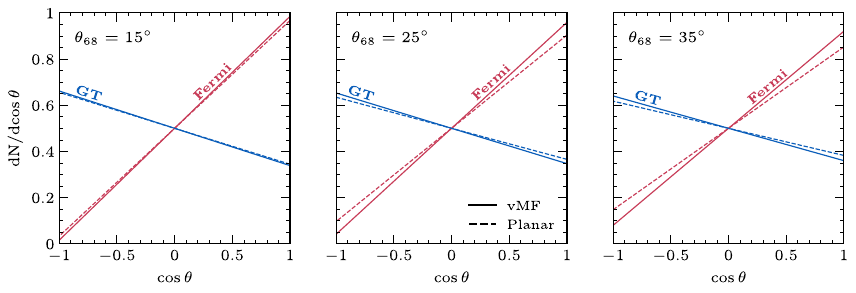}
    \caption{Comparison between von Mises-Fisher and planar approximation smearing over the Fermi and Gamow-Teller angular distributions. \emph{For typical LArTPCs angular resolutions, there is a notable effect between the correct and approximated smearing.}}
    \label{fig:vMF_vs_planar}
\end{figure}

\Cref{fig:vMF_vs_planar} illustrates the mentioned discrepancy between the von Mises-Fisher distribution and the planar approximation treatment. In this analysis, it is fundamental to account for the proper smearing function on a sphere if the final configuration has an angular resolution $\gtrsim$25$\degree$.
We reemphasize that this treatment involves only analytic formulas and does not introduce extra computational complications. Thus, even if the planar approximation works well, there is no reason not to use the proper description.

\subsection{Likelihood}
\label{sec:App_Chi2}

In order to study the potential sensitivity for cross section extraction using solar $^{8}\mathrm{B}$ neutrinos we neglect the tiny uncertainties of the initial flux and the oscillation parameters involved. Each cross section transition depends on two quantities: the transition energy, which is extremely well measured and therefore assumed to be perfectly known, and the transition strength, which in principle is the quantity we want to extract in order to construct the correct cross section. Since we already have indirect measurements in Ref.~\cite{Bhattacharya:2009zz}, instead of a pure measurement of the transition strengths we rather focus on deviations from the indirect measurement.

In the following, neglecting all the mentioned sources of uncertainties, the parameters we aim to extract in all analyses are the normalizations mentioned in the main text, described collectively by $\vec{n}$ due to the variability on the number of parameters used per method. In order to study the sensitivity to each individual component of $\vec{n}$, we perform a binned maximum likelihood analysis of the event rate at DUNE, and use the likelihood ratio as our test statistics. The total log-likelihood-ratio reads as
\begin{equation}
\label{eq:chi2}
\Delta \chi^2 (\vec{n}) = 2  \sum_{k} \left[ N_{\mathrm{ev},\,k}(\vec{n}) - N_{\mathrm{ev},\,k}^{(p,n)} + N_{\mathrm{ev},\,k}^{(p,n)}\ln\left(\frac{N_{\mathrm{ev},\,k}^{ (p,n)}}{N_{\mathrm{ev},\,k}(\vec{n})}\right)\right] \,,
\end{equation}
where $N_{\mathrm{ev},\,k}(\vec{n})$ is the expected number of events in a particular energy and angular bin, $k$, assuming a cross section described by the particular prescription chosen for each case and $\vec{n}$ for the normalization values. The expected number of events on the bin $k$ for the transition energies and strengths in \Cref{tab:E_and_strengths} is $N_{\mathrm{ev},\,k}^{(p,n)}$, which will serve as our benchmark to test the sensitivity. For each case, we assume that the log-likelihood-ratio in \cref{eq:chi2} follows a $\chi^2$ distribution with $\dim(\vec{n})$ degrees of freedom.

\newpage
\section{The $\boldsymbol{i=1}$ and $\boldsymbol{i=2}$ strong degeneracy}\label{sec:App_12deg}

The most significant limitation of our pessimistic analysis is the inability to reconstruct the first transition. The strong degeneracy between the $i=1$ and $i=2$ transitions (\Cref{tab:E_and_strengths}) requires us to choose between resolving the two dominant GT transitions or achieving a precision extraction of the cross section in the low-MeV region.

To show this, we examine the most optimistic scenario for decoupling these transitions, specifically the case in which the cross section consists exclusively of these two components
\begin{equation}\label{eq:deg1and2}
    \tilde{\sigma}_\mathrm{Ar} = n_{1}\sigma_1+n_2\sigma_2\,.
 \end{equation}    

\begin{figure}[h!]
    \centering    \includegraphics[width=0.56\columnwidth]{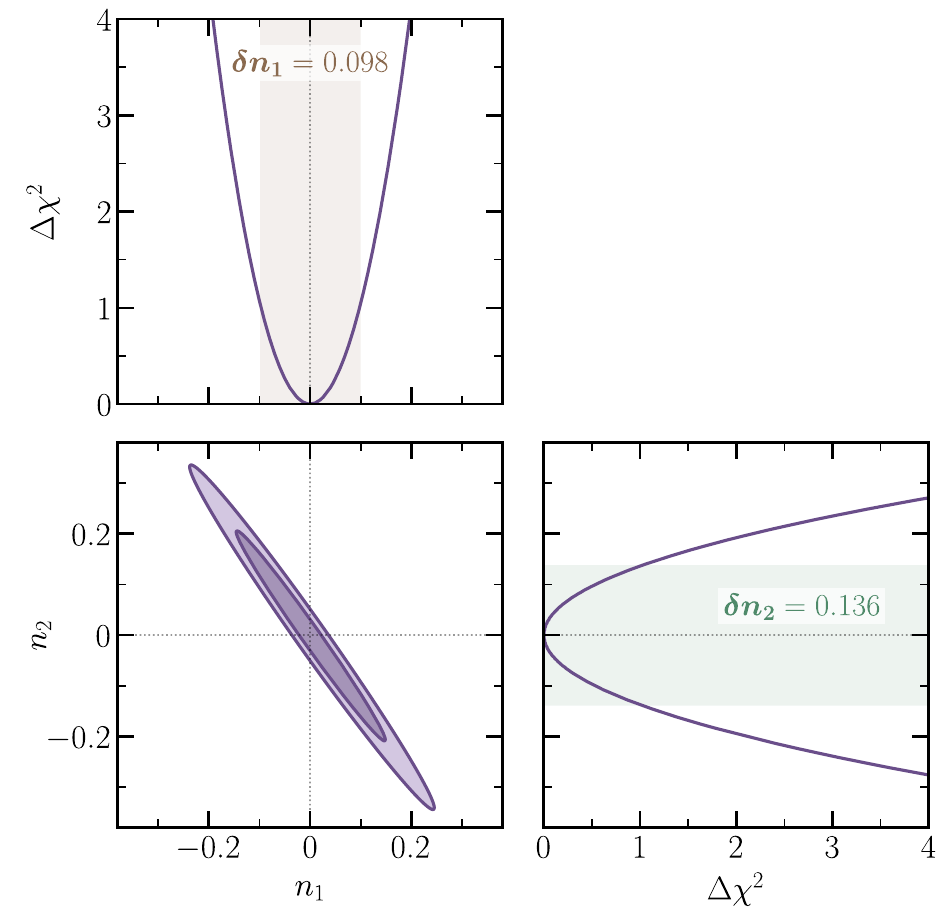}
    \caption{Degeneracy and uncertainty in the normalization of the two transitions described in \cref{eq:deg1and2}. If one attempts to disentangle these transitions, it becomes impossible to reconstruct the cross section with high precision. Consequently, an effective grouping of transitions is required to extract the cross section with low uncertainty.}
    \label{fig:App_deg1and2}
\end{figure}

\Cref{fig:App_deg1and2} shows an unrealistic optimistic extraction of the first and second transitions, obtaining a precision on each transition of 
\begin{align}
    \delta n_{1} &=  9.8\,\%\,, \nonumber\\
    \delta n_{2} &= 13.6\,\% \,, \nonumber 
\end{align}
leading to a minimum uncertainty of $\sim$10\% for the cross section extraction. However, this poor situation is illusory. In the realistic case where additional subdominant transitions exist, the sensitivity worsens, rendering it unfeasible to improve upon the current 10\% uncertainty from indirect measurements if one wants to resolve these two states. 

The greater the number of GT transitions included, the poorer the expected sensitivity. Fundamentally, these transitions differ only in their energy spectra. However, identifying an independent GT sample is challenging unless a transition possesses a distinct energy and sufficient strength. This is not the case in reality if one accounts for the indirect measurements expected values. Consequently, the parameter to which our analysis is most sensitive is $\sigma_1+\sigma_2$, which effectively encapsulates the dominant event count of the GT sector of the cross section. This sum can be determined at the $\sim$2\% level, as shown in the main text.

\newpage

\section{Principal Component Analysis of the cross section}\label{sec:App_PCA}

It is natural to assume that calibrating the cross section requires determining the strengths of the 15 individual transitions. However, high-precision knowledge of individual strengths is not strictly required to characterize the cross section shape. This is partly due to the extremely well-measured energies of each transition, $\Delta E_i$~\cite{Bhattacharya:1998hc, Bhattacharya:2009zz}, but also because of the existence of a hierarchical, highly degenerate distribution in which a small number of strengths dominate the full energy range.

Initial estimates of the strength values, based on measurements in Refs.~\cite{Bhattacharya:1998hc, Bhattacharya:2009zz}, combined with energy and angular distributions, allow for the extraction of the cross section in DUNE by focusing on $^8\mathrm{B}$ solar neutrinos. However, due to the aforementioned degeneracies between different transitions, identifying individual strengths is challenging and leads to large, artificial uncertainties. The Principal Component Analysis (PCA) is particularly useful in this context. It identifies the effective number of degrees of freedom available for testing and provides linear combinations of strengths ordered by their variance. Although these combinations may lack immediate interpretability, they can be readily mapped back to the 15 constituent strengths. Furthermore, rather than using the linear combinations from PCA directly, it is sufficient to use the results to guide the analysis and maintain intuitive variables.

The methodology of PCA is as follows:
\begin{itemize}
    \item We start by placing all information available in a matrix $\mathcal{M}$ of dimension ($n_\mathrm{bins}$, $n_\mathrm{parameters}$). In our case this corresponds to the expected number of events per bin, for each of the 15 transitions.
    \item Data needs to be standardized per variable, centering each one around its mean value, generating the matrix
    \begin{equation}\label{eq:Zdata}
        \mathcal{Z} = \mathcal{M} - \overline{\mathcal{M}}\,, 
    \end{equation}
    where $\overline{\mathcal{M}}$ denotes a matrix that has the mean value of the number of events for each transition. This corresponds to a transformation shifting the origin from zero to the mean value of events per transition.

    Some analyses define the standardized variable as $\tilde{\mathcal{Z}} = (\mathcal{M} -\overline{\mathcal{M}}) / \sigma$. This approach would effectively discard the prior information established by existing indirect measurements~\cite{ICARUS:1998nzl, Bhattacharya:1998hc, Bhattacharya:2009zz}. To maintain consistency with them, we employ the definition specified in \cref{eq:Zdata} for our PCA framework.
    
    \item We compute the covariance matrix
    \begin{equation}
        \mathcal{C} = \frac{\mathcal{Z}^T\mathcal{Z}}{N-1}\,, 
    \end{equation}
    where $N$ is the initial dimensionality of our problem, in this case $N=15$.
    \item We extract the eigenvalues ($\lambda_i$) and eigenvectors ($\vec{u}_i$) of the covariance matrix by solving
    \begin{equation}
        |\mathcal{C} -\lambda I | = 0 \,.
    \end{equation}
    \item We order the eigenvalues from higher absolute value to lower, and compute the cumulative variance, defined as
    \begin{equation}
        V_i = \frac{\sum_{k=1}^i \lambda_k}{\sum_{k=1}^N \lambda_k} \,.
    \end{equation}
    The eigenvalues' meaning is that of the variance of the direction of its corresponding eigenvector. Higher variance means higher sensitivity to that particular direction from our data.
    
    \item We keep only states until we surpass the $V_{i_\mathrm{max}}\sim0.95$ threshold. This ensures avoiding strong degeneracies between the states, effectively reducing the dimensionality of the problem without significant loss of information. In our case, $i_\mathrm{max}=2$ when accounting for both energy and angular spectra. Therefore, the PCA method reduced our dimensionality from 15 to 2, leading to a sufficient description of the cross section with
    \begin{equation}
        \sigma_\mathrm{Ar} = \left(\vec{\mathbf{1}}+n_1\vec{v}_1+n_2\vec{v}_2 \right)\cdot\vec{\sigma}\,,
    \end{equation}
    where the two new parameters describing the cross section are $n_1$ and $n_2$.
\end{itemize}

In the following, we describe the PCA characteristics for each extraction method that utilizes this approach. In this paper, we use the python library \texttt{sklearn}~\cite{scikit-learn} to perform the PCA analysis of the cross section.

\subsection{Cross section extraction with 2 PCA vectors}
\label{sec:app_2pca}

The most general scenario considered is the application of PCA to the pessimistic case, exploiting the full dimensionality of the available information, including detected energy and angular distributions. This analysis is fully guided by the PCA framework to maximize extraction sensitivity with the lowest dimensionality possible.

\begin{figure}[h!]
    \centering    \includegraphics[width=\columnwidth]{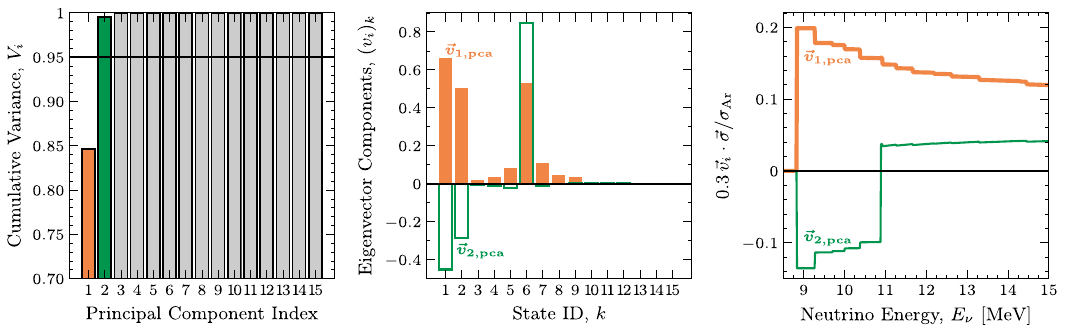}
    \vspace*{-0.5cm}
    \caption{PCA analysis of the 15 transitions cross section.}
    \label{fig:App_PCA_2PCA}
\end{figure}

\Cref{fig:App_PCA_2PCA} shows the results of the PCA analysis on the cross section transitions. The first panel shows the cumulative relative variance for the 15 vectors ordered from higher to lower variance. In this case, the dimensionality of the problem is reduced to 2. The associated eigenvectors for these two states are shown in the second panel of the figure. Leaving aside the relative weights and additional contributions from other states, the PCA relevant vectors could be seen mainly as $v_\mathrm{1,pca} \simeq \sigma_1+\sigma_2+\sigma_6$ and $v_\mathrm{2,pca} \simeq-\sigma_1-\sigma_2+\sigma_6$. This further validates that the dominant transitions of the analysis are $\sigma_1$, $\sigma_2$ and $\sigma_6$. The last panel of the figure illustrates the direct impact on the cross section of a 0.3 contribution of each vector, by showing the relative deviation with respect to the standard case.

\vspace{-0.5cm}
\subsection{Cross section extraction with Fermi + 1 PCA}
\label{sec:app_Fermi+1PCA}

Guided by \Cref{fig:App_PCA_2PCA} and the distinct angular distribution of the $i=6$ transition relative to the others, there is a clear physical basis for isolating this specific state. By doing so, we can perform a direct measurement of the $i=6$ state's transition strength. This allows for comparison with the theory value for superallowed Fermi transitions.

\begin{figure}[b]
    \centering    \includegraphics[width=\columnwidth]{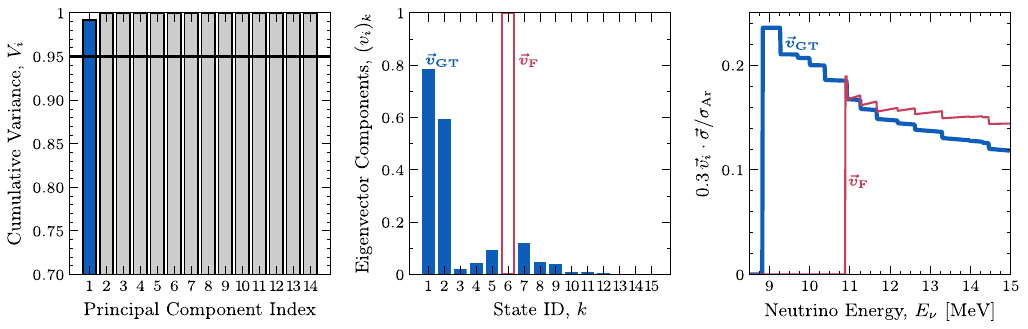}
    \vspace*{-0.5cm}
    \caption{PCA analysis of the Gamow-Teller transitions.}
    \label{fig:App_PCA_Fermi+1PCA}
\end{figure}

\Cref{fig:App_PCA_Fermi+1PCA} shows the results of the PCA analysis of the Gamow-Teller states, and includes the Fermi state that we isolated by hand for comparison. On the first panel one sees that there is only one degree of freedom that our data is able to constrain within Gamow-Teller transitions. Indeed, this state is qualitatively similar to the 2 PCA one when removing the $i = 6$ state. This is visible on the second panel of the figure. Finally, in the third panel we can see the cross section changes due to the inclusion of the Fermi and PCA-GT states.

\clearpage
\twocolumngrid

\bibliographystyle{JHEP.bst}
\bibliography{biblio}

\end{document}